%% file: main.tex
\documentclass[manuscript]{acmart}

\AtBeginDocument{%
  }

\setcopyright{acmlicensed}
\copyrightyear{2025}
\acmYear{2025}
\acmDOI{XXXXXXX.XXXXXXX}

\usepackage{graphics}
\usepackage{tabularx}
\usepackage{xspace}
\usepackage{xcolor}
\usepackage{multirow}
\usepackage{booktabs}

\usepackage{amsmath,amssymb,amsfonts}
\usepackage{textcomp}
\usepackage{url}
\usepackage[colorinlistoftodos,prependcaption,textsize=tiny]{todonotes}
\usepackage{enumitem}
\usepackage{graphicx}
\usepackage{adjustbox} 
\usepackage{tabularx}  
\usepackage{tcolorbox}  
\usepackage{lipsum}     
\usepackage{hyperref}  
\usepackage{subfigure}
\usepackage{booktabs}
\usepackage{balance}
\usepackage{bm}

\usepackage{algorithm}
\usepackage{algorithmic}
\usepackage{multirow}
\usepackage{soul}

\newcommand{\etal}{{\em et al.}\xspace}
\newcommand{\ie}{{\em i.e.},\xspace}
\newcommand{\eg}{{\em e.g.},\xspace}

\newcommand{\methodname}{{AL-Bench}\xspace}

\begin{document}
\author{Boyin Tan}
\email{BoyinTan@link.cuhk.edu.cn}
\affiliation{
  \institution{The Chinese University of Hong Kong, Shenzhen}
  \country{China}
}

\author{Junjielong Xu}
\email{junjielongxu@link.cuhk.edu.cn}
\authornote{Junjielong Xu and Pinjia He are corresponding authors.}
\affiliation{
  \institution{The Chinese University of Hong Kong, Shenzhen}
  \country{China}
}

\author{Zhouruixing Zhu}
\email{zhouruixingzhu@link.cuhk.edu.cn}
\affiliation{
  \institution{The Chinese University of Hong Kong, Shenzhen}
  \country{China}
}

\author{Pinjia He}
\email{hepinjia@cuhk.edu.cn}
\authornotemark[1]
\affiliation{
  \institution{The Chinese University of Hong Kong, Shenzhen}
  \country{China}
}

\input{sections/001_abstract}
\title{\textit{\methodname}: A Benchmark for Automatic Logging}



\maketitle
\tcbset{
    colback=gray!10,    
    colframe=black,     
    boxrule=0.5mm,      
    arc=3mm,            
    auto outer arc,
    width=\textwidth,   
    left=5pt,          
    right=5pt,          
    boxsep=5pt,         
    before=\vskip10pt,  
    after=\vskip10pt,   
    center    
}

\input{sections/002_introduction}

\input{sections/003_motivation}

\input{sections/004_method}

\input{sections/005_experiments}

\input{sections/006_conclusion}

\bibliographystyle{ACM-Reference-Format}
\bibliography{ref}

\end{document}

%% file: sections/001_abstract.tex
\begin{abstract}
Logging, the practice of inserting log statements into source code, is critical for improving software reliability. 
Recently, language model-based techniques have been developed to automate log statement generation based on input code. 
While these tools show promising results in prior studies, the fairness of their {results} comparisons is not guaranteed due to the use of ad hoc datasets.
In addition, {existing evaluation approaches exclusively dependent on code similarity metrics fail to capture the impact of code diff on runtime logging behavior, as minor code modifications can induce program uncompilable and substantial discrepancies in log output semantics.}
To enhance the consistency and reproducibility of logging evaluation, we introduce \methodname, a comprehensive benchmark designed specifically for automatic logging tools. \methodname includes a large-scale, high-quality, diverse dataset collected from 10 widely recognized projects with varying logging requirements. Moreover, it introduces a novel dynamic evaluation methodology to provide a run-time perspective of logging quality in addition to the traditional static evaluation at source code level.
Specifically, \methodname not only evaluates the similarity between the oracle and predicted log statements in source code, but also evaluates the difference between the log files printed by both log statements {during runtime}.
\methodname reveals significant limitations in existing static evaluation, as all logging tools show average accuracy drops of 37.49\%, 23.43\%, and 15.80\% in predicting log position, level, and message compared to their reported results.
Furthermore, with dynamic evaluation, \methodname reveals that 20.1\%-83.6\% of these generated log statements are unable to compile. 
Moreover, the best-performing tool achieves only 21.32\% cosine similarity between the log files of the oracle and generated log statements.
These results underscore substantial opportunities to advance the development of automatic logging tools.
We believe this work establishes a foundational step in furthering this research direction.

\end{abstract}



%% file: sections/002_introduction.tex
\section{introduction}
\label{sec:intro}
As software grows in size and complexity, logging has become increasingly essential to ensuring software reliability~\cite{he2021survey, Chen2021ASO}. 
Logging means writing log statements into the source code, which generate runtime logs that record valuable information for a range of downstream tasks such as anomaly detection \cite{Zhang2019RobustLA, Lou2010MiningIF, He2016ExperienceRS, Du2017DeepLogAD, Zhang2019RobustLA}, fault diagnosis \cite{Zou2016UiLogIL,xu2023hue, xu2024divlog}, root cause analysis \cite{Amar2019MiningHT, He2018IdentifyingIS, Lin2016LogCB,xu2025openrca}, and program analysis \cite{Ding2015Log2AC, Shang2013AssistingDO, xu2024aligning}. 
The effectiveness of these downstream tasks heavily relies on the quality of the software logs~\cite{He2018CharacterizingTN}. Therefore, appropriate logging is essential to capture critical behaviors during software operation~\cite{YuanUsenixA1}.
\input{insert/logging_example}

To help software developers implement effective and efficient logging practices, researchers have been focusing on exploring methodologies for automatic logging in recent years.
As illustrated in Figure \ref{fig:logging_example}, automatic logging typically includes three steps: (1) determining the position, (2) deciding the verbosity level, and (3) specifying the message to be recorded.
To address the automatic logging tasks, numerous logging methods have been proposed thanks to the advanced development of the Large Language Model (LLM).
For example, \textit{LANCE}~\cite{Mastropaolo2022UsingDL} was introduced as the first end-to-end logging method, seamlessly integrating the recommendation of log statement position, verbosity level, and message content.
Building on this foundation,
\textit{UniLog} \cite{Xu2024UniLogAL} employed a warm-up and in-context learning (ICL) strategy to enhance performance. 
\textit{FastLog} \cite{xie2024fastlog} improved the generation efficiency while maintaining precision. 
\textit{LEONID} \cite{Mastropaolo2023LogSG}, based on \textit{LANCE}, combined with deep learning and information retrieval technologies to enhance performance. \textit{SCLogger}~\cite{Li2024GoSC} adapted static analysis to extend the context for the code snippet.


However, current logging research faces a significant challenge: \textbf{The absence of a standardized evaluation dataset and methodology hinders fair comparisons between different logging methods}. This issue manifests in two key aspects:
(1) The test sets of different logging methods are often derived from their own training data, leading to inconsistent experimental performance and hindering fair comparisons.
(2) The evaluation metrics used for different logging methods are not fully consistent, resulting in varying advantages across different dimensions.
Given these challenges, we argue that \textit{establishing a comprehensive logging benchmark is essential}. However, it is infeasible to naively construct a benchmark via \textit{ensembling} existing evaluation data and metrics due to two major limitations of existing evaluation datasets and methods:

\textbf{First, existing evaluation datasets lack the quality required for reliable assessment.}
For example, 18.45\% cases (2218/12020) in LANCE's test set and 30.36\% cases (2197/7237) in FastLog's test set involve numerous bad patterns of log messages. 
As illustrated in Figure~\ref{fig:low_quality} these anomalies predominantly fall into the following types: (a) duplicated content, (b) empty string, (c) duplicated empty and special characters, (d) contents that mismatch the log levels, and (e) explicit type casting.
These issues distort evaluation outcomes, as tools producing such flawed log statements tend to appear better on metrics despite having poor logging quality.
Moreover, these datasets are restricted to cases with very short contents ($\leq$ 512 tokens), a limitation imposed by their use of foundation models (e.g., T5, PLBART), which support a maximum input of 512 tokens. This constraint not only reflects a compromise to accommodate their models but also inhibits the evaluation of realistic, function-level logging capabilities, as many industrial functions exceed this token limit. Consequently, the datasets fail to meet the requirements for assessing logging quality in real-world scenarios.
\textbf{Second, current evaluation methods lack a fine-grained view for comprehensive assessment.}
Existing evaluations primarily focus on static code similarity between candidate log statements generated by logging methods and oracle log statements written by developers.
However, the code compilability is entirely neglected and can not be measured by static evaluation.
Our preliminary analysis revealed that 20\%-80\% of log statements generated by current methods fail to compile, a figure that has not been previously reported.
Furthermore, such static similarity fails to capture logging performance during program execution.
For instance, placing a log statement inside or outside a \textit{loop} may differ by only one line in code but result in vastly different numbers of log entries at runtime. Conversely, moving a log statement above or below a \textit{comment} may have no effect on the runtime logs despite a one-line difference in the code. Such runtime differences are overlooked in static evaluations.
These limitations highlight fundamental flaws in the current evaluation approach.
\input{insert/low_quality_data_example}
To address these two challenges, our key insights are: (1) \textit{\textbf{Log statements from the most widely used and frequently updated repositories tend to be of higher quality with fewer bad logging patterns.}} (2) \textit{\textbf{Log files printed in runtime can serve as a proxy for evaluating log statements quality with a more comprehensive view of program logging performance.}}
Based on these insights, we introduce \methodname, the first unified logging benchmark involving a large-scale evaluation dataset and a comprehensive evaluation method for assessing both the static log statements and runtime log files printed via code execution.
Our dataset comprises 21,804 instances mined from the 10 most popular and actively maintained GitHub projects~\cite{GitHub}, each with at least 10,000 stars, and 500 log-related issues spanning diverse domains and application scenarios.
Furthermore, \methodname introduces \textit{dynamic evaluation} in addition to traditional static evaluation for log statements.
Specifically, it not only applies a suite of static metrics to quantify the quality of generated log statements but also reintegrates them into real project code, compiles, and executes them.
This approach enables a realistic assessment beyond the traditional static approach, highlighting major limitations in existing logging methods:  even the best-performing logging method fails to compile in 20.1\% of cases, and the logs printed in runtime show only 21.32\% cosine similarity to the oracle logs.
By providing a unified, standardized, and comprehensive evaluation dataset and methodology, we believe \methodname establishes a foundational step toward advancing research in automatic logging.

To sum up, our contributions are shown as follows:
\begin{enumerate}
    \item We collected a high-quality, diverse, and large-scale dataset comprising 21,804 instances from 10 popular, high-quality GitHub~\cite{GitHub} projects, spanning various domains with differing logging requirements.
    \item We propose a dynamic evaluation approach that reintegrates the generated log statements into projects to evaluate their printed runtime logs as a more fine-grained proxy for illustrating logging performance.
    \item We conducted a comprehensive evaluation of popular automatic logging tools and revealed the key limitations based on the analysis of the evaluation results. 
    \item All the data and code for \methodname are publicly available\footnote{\url{https://github.com/shuaijiumei/logging-benchmark-scripts}}, providing valuable resources for both developers and researchers to advance the field of automatic logging.
\end{enumerate}


%% file: insert/logging_example.tex
\begin{figure*}[ht]
    \centering
    \includegraphics[width=0.9\textwidth]{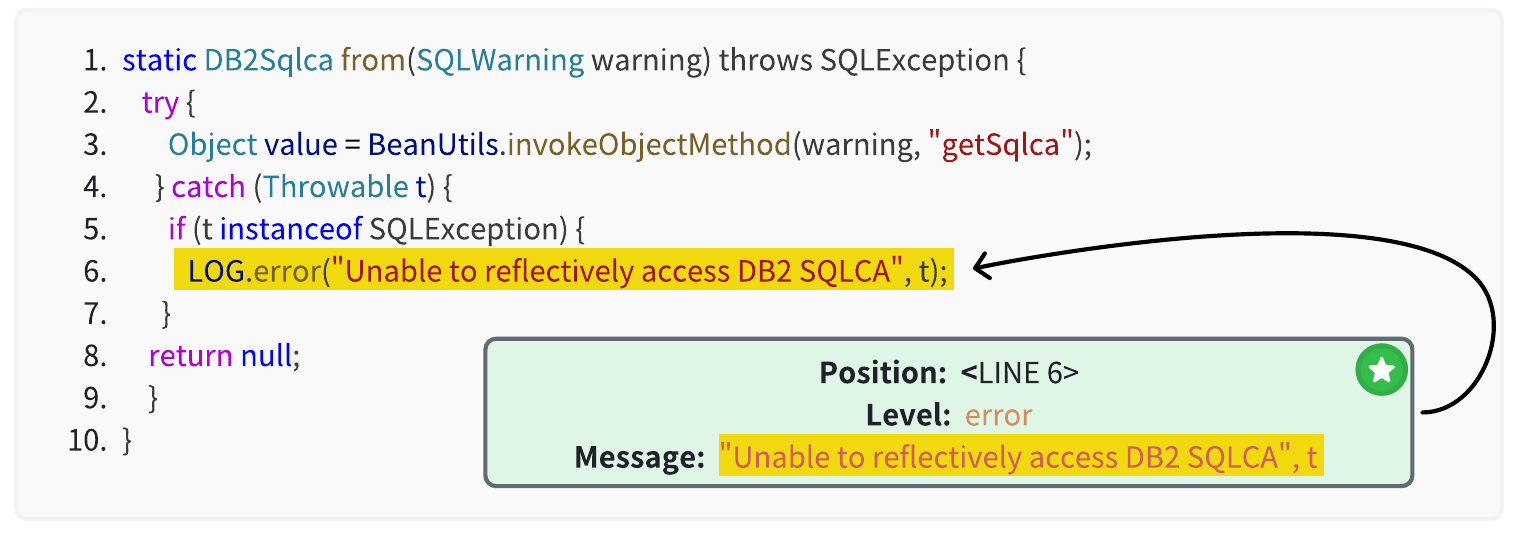}
    \caption{An example of logging statement generation. Logging statement generation can be separated as three parts: determining the position, selecting the level, and specifying the message.}
    \label{fig:logging_example}
\end{figure*}

%% file: insert/low_quality_data_example.tex
    \begin{figure}[t]
    \centering
    \includegraphics[width=0.9\linewidth]{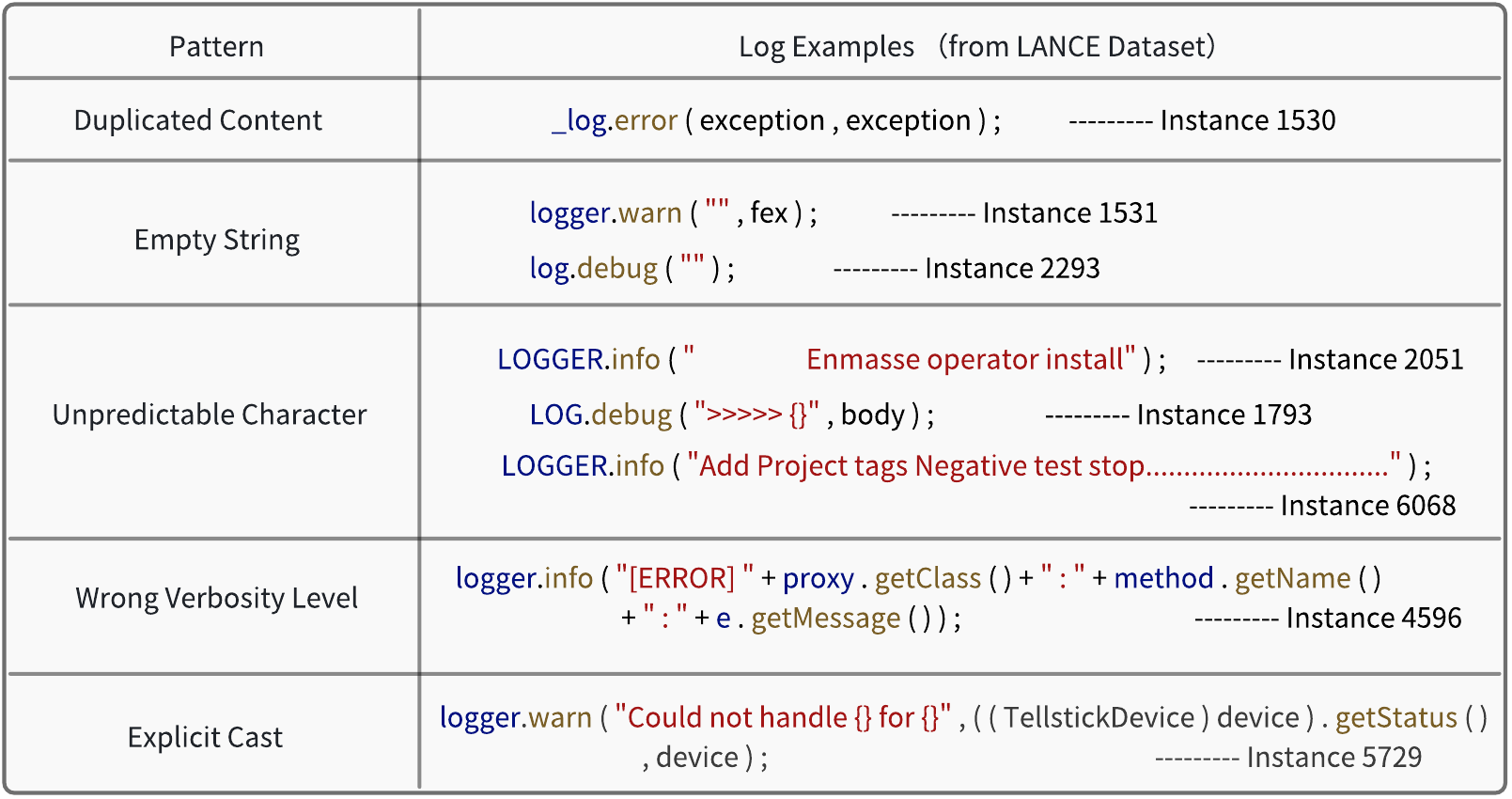}
    \caption{Bad patterns in existing datasets (using instances from LANCE dataset as examples): (1) \textbf{Duplicated Variable} records the same information multiple times, costing redundant overhead for both printing and recording logs in runtime. (2) \textbf{Empty String} provides insufficient context, hindering the effectiveness of debugging through printed logs. (3) \textbf{Unpredictable Character} contains numerous meaningless special tokens, making both the logging methods hard to predict and their printed logs difficult to parse for downstream analysis~\cite {gojko2006logging}. (4) \textbf{Wrong Verbosity Level} often misleads developers for debugging and fault localization~\cite{Chen2017CharacterizingAD}. (5) \textbf{Explicit Cast} couple logs to variable type casting, might cause runtime type conversion errors and system crash~\cite{Chen2017CharacterizingAD}.
    }
    \label{fig:low_quality}
    \end{figure}

%% file: sections/003_motivation.tex
\section{Background}
\label{sec:moti}

\subsection{Related Work of Automatic Logging}
Logging, the process of generating informative log messages with appropriate verbosity levels at strategically placed locations within code, has long been recognized as a critical challenge in software engineering~\cite{he2021survey, Chen2019ExtractingAS, Chen2021ASO, Chen2021ExperienceRD}. Over the years, substantial research efforts have aimed to support developers in crafting more effective logging statements, which in turn enhance software maintenance and testing~\cite{Jia2018SMARTLOGPE, Zhao2017Log20FA, Zhu2015LearningTL}. Early studies in this domain often addressed isolated subproblems, typically operating under stringent assumptions that limit the applicability of their findings in real-world scenarios. For example, Li \etal~\cite{Li2021DeepLVSL} proposed \textit{DeepLV} to predict the appropriate logging level by taking surrounding code features into a neural network. Liu \etal~\cite{Liu2022TeLLLL} proposed \textit{Tell} to further adapted flow graphs to help the suggestions of verbosity levels. Zhu \etal~\cite{Zhu2015LearningTL} proposed \textit{LogAdvisor} and Yao \etal~\cite{Yao2019Log4PerfSA} proposed \textit{Log4Perf} to assist developers to add new log statements in a specific position. Ding \etal proposed \textit{LoGenText}~\cite{Ding2022LoGenTextAG} and \textit{LoGenText-Plus}~\cite{Ding2023LoGenTextPlusIN} to advise developers what should be logged, and Liu \etal \cite{Liu2021WhichVS} proposed tools for deciding which variables should be logged. However, none of them can generate a complete log statement. 

Recently, with the advanced capabilities of large language models (LLMs)~\cite{Floridi2020GPT3IN, raffel2020exploring}, numerous LLM-based logging methods have been proposed. 
Specifically, \textit{LANCE}~\cite{Mastropaolo2022UsingDL} was introduced as the first end-to-end logging method, seamlessly integrating the recommendation of log statement position, verbosity level, and message content.
Building on this foundation, \textit{UniLog} \cite{Xu2024UniLogAL} employed a warm-up and in-context learning (ICL) strategy to enhance performance. 
\textit{FastLog} \cite{xie2024fastlog} improved the generation efficiency while maintaining precision. 
\textit{LEONID} \cite{Mastropaolo2023LogSG}, based on \textit{LANCE}, combined with deep learning and information retrieval technologies to enhance performance. \textit{SCLogger}~\cite{Li2024GoSC} adapted static analysis to extend the context for the code snippet.

    

\subsection{Philosophy of Logging Evaluation}\label{sec:phy}


Logging quality evaluation has always been a critical yet challenging problem in automatic logging~\cite{Bogatinovski2022QuLogDA, Li2019DLFinderCA, Chen2017CharacterizingAD, Chen2021ASO, he2021survey}. In an ideal scenario, high-quality logs should accurately capture key runtime state information of a software system while avoiding excessive logging that may lead to unnecessary consumption of computational and storage resources. Intuitively, the most direct approach to assessing logging quality is to evaluate how well the logs support specific \textbf{downstream tasks}~\cite{YuanUsenixA1}. However, using downstream tasks as proxies to measure general-purpose logging quality in a scalable way is both theoretically and practically problematic for several reasons:

\begin{enumerate}
    \item First, it is very challenging to cover all potential downstream tasks for logs exhaustively. There are countless log-based downstream tasks: beyond the tasks with clear task formulations (e.g., debugging, anomaly detection, root cause analysis), logs are also used for a wide range of tasks that lack a universal formulation (e.g., recording sequences of runtime events to monitor system status), often driven by the unique needs of each system’s developer~\cite{Zhang2019RobustLA, xu2024divlog, Amar2019MiningHT}. Even if we only choose a fixed set of tasks with clear formulations to evaluate logging quality, these proxies can only assess \textit{task-specific aspects} rather than provide a comprehensive measure of \textit{overall} logging quality. This limitation makes them unreliable indicators of \textit{general-purpose} logging quality.
    \item  Second, it is not practical to build testbeds for these downstream tasks across all software systems. While it might be theoretically possible to create a testbed for evaluating logging quality in common tasks like anomaly detection, assessing tasks like whether logs capture key information for event monitoring needs is challenging, as it depends heavily on subjective developer requirements. Moreover, even creating fault datasets to support log-based anomaly detection is also challenging since different software systems have unique fault characteristics, making scalable fault injection across diverse environments infeasible~\cite{zhao_identifying_2021}.
    \item Third, the trade-off between logging effectiveness and efficiency is another unmeasurable aspect of logging quality. Massive logging offers detailed runtime data for fault diagnosis but increases storage costs and resource usage, while sparse logging reduces expenses but risks missing critical information. System designers balance these tradeoffs primarily through operational experience from maintaining and debugging specific systems. However, such experiences are inherently context-dependent, varying across systems due to divergent functional, architectural, and operational priorities~\cite{rong2023developers}. Consequently, it remains challenging to establish a universal trade-off mechanism for assessing the appropriateness of logging strategies for each specific software.
\end{enumerate}

As an alternative, logging research often adopts a \textbf{similarity-based} approach to evaluate logging quality.
The core idea is that, since assessing log quality via downstream tasks is challenging, we can use high-quality log statements as references to evaluate new ones.
Thus, by constructing benchmarks, we can check if a logging method generates log statements similar to these reference ones, making evaluation scalable with log statements from open-source software.

However, this similarity-based approach does have limitations.
It may not accurately assess log statements of \textit{higher quality} than the reference, as they could receive lower similarity scores simply for not matching closely enough. To mitigate this, reference log statements must be of sufficiently high quality, which poses a challenge in data collection. We argue that log statements from the most \textit{widely-used}, \textit{time-tested} projects are the best reference candidates. Additionally, current evaluations focus solely on static code, but small changes in code can lead to significant differences in log content. Therefore, evaluating both static code and the generated log files provides a more comprehensive measure of logging quality. These insights inform the development of our first logging benchmark: \methodname.

%% file: sections/004_method.tex
\section{\methodname}
\label{sec:method}
\methodname is a novel benchmark designed to evaluate automatic logging tools in codebases systematically. The benchmark comprises 21,804 code snippets and 39,600 log statements extracted from 10 most popular open-source repositories, selected for their active maintenance, diverse application domains, and representation of modern software engineering practices. \methodname employs a dual evaluation framework: \textbf{Static Evaluation} evaluates the \textit{similarity of log statements} (\eg \textit{position}, \textit{level}, \textit{variable}, \textit{message}) between predicted and reference log statements, while \textbf{Dynamic Evaluation} assesses code \textit{compileability} and the \textit{similarity of logs} printed by those log statements during runtime execution. Both evaluation methods will be described later in this section.



\subsection{Dataset Construction}
\label{sec:method:static_dataset_building} 
As discussed in Sec.~\ref{sec:intro}, the quality of the evaluation dataset is crucial for assessing the performance of tools. 
Building on empirical evidence that GitHub repository stars correlate with code quality~\cite{Jiang2024ASO}, we determine dataset inclusion criteria for log statement analysis by requiring repositories to have $\geq$ 10,000 stars and $\geq$ 500 log-related issues to ensure community validation and logging relevance. Candidate repositories are then ranked by total log statement count to prioritize those with extensive logging practices.
From the top of this ranked list, we manually curate a final dataset of 10 repositories to ensure that the set covers different priorities in the logging effectiveness-efficiency trade-off, including performance-critical systems (requiring minimal logs) and trace-intensive applications (requiring detailed logs). After that, we ensure that all of these repositories are actively maintained, with frequent commits (>50/month), a high issue resolution rate (>70\%), and recent updates (within the past 6 months). This dual-phase approach balances the quantitative scale with domain representativeness.

As shown in Table~\ref{table:dataset}, our final dataset includes projects with a total of 21,804 code snippets and 39,600 log statements, covering a wide range of logging needs and practices.
The dataset spans multiple domains, including database management, task scheduling, distributed storage, messaging systems, and IoT platforms.
These repositories have \textit{\textbf{distinct logging objectives (targeted downstream tasks)}} based on the specific needs of their respective scenarios and exhibit \textit{\textbf{unique effectiveness-efficiency trade-offs (more logging vs less logging)}}, as shown in Sec.~\ref{sec:phy}.
For example, database systems like DBeaver minimize logging overhead to sustain performance, while task schedulers such as DolphinScheduler emphasize dependency-tracing logs for runtime monitoring. Distributed systems (e.g., Hadoop) enforce rigorous logging for fault tolerance and distributed coordination, contrasting with messaging platforms like Kafka, which track message flows to ensure delivery reliability. IoT solutions (e.g., ThingsBoard) leverage logs for real-time device connectivity oversight, whereas security-focused systems like Keycloak prioritize log anonymization to safeguard sensitive data.

These diverse logging requirements not only demonstrate the heterogeneity and inherent contradictions in logging practices (e.g., performance-sensitive scenarios demand minimal logging overhead, while debugging scenarios necessitate comprehensive event records) but also substantiate the impossibility of establishing universal standards for assessing log code quality.
By documenting real-world logging practices from community-vetted, widely-used projects across domains, we establish a foundation for evaluating logging strategies in their native environments. This domain-rooted dataset supports comprehensive analysis of automated logging techniques across diverse operational scenarios.
    

 \input{insert/table_dataset_projects}

\subsection{Static Evaluation}
\label{sec:statci_evaluation}
\subsubsection{Task Formulation}
We structure each entry in our dataset as a tuple \(<Code_{w/o~ LogStmt}, LogPos, LogStmt>\). As illustrated in Figure~\ref{fig:static_evaluation_task_formulation},  \(Code_{w/o~ LogStmt}\) represents the input code context with one log statement deliberately removed, \(LogPos\) indicates the ground truth position for the missing log statement, and \(LogStmt\) represents the exact log statement to be predicted.
 This process yielded 39,600 high-quality instances, providing a robust foundation for evaluating log generation models.

 \subsubsection{Metrics}
 \label{sec:method:metrics}


We introduce six static evaluation metrics to assess the similarity between the generated and reference log statements.

\textit{\textbf{Metric 1: Position Accuracy (PA):}} PA focuses on the precise line of log statements within the source code. The correct placement of log statements helps accurately trace the execution flow and diagnose issues. For Position Accuracy, we rigorously compare the predicted positions of log statements with their actual positions in the source code. This metric is calculated by taking the number of correctly positioned log statements \(P_c\) and dividing it by the total number of log statements \(N_a\) to obtain the accuracy value: \(PA = \frac{P_c}{N_a}\).

\textit{\textbf{Metric 2: Level Accuracy (LA):}}  
LA evaluates the exact match between predicted and reference log levels, which are essential for prioritizing operational events in DevOps pipelines. Common log levels carry distinct semantic implications, like "\textit{info}," referring to the normal information of runtime behavior.  "\textit{Warning}" indicates potential problems that might not immediately cause disruption but could lead to future issues if not resolved. "\textit{Error}" refers to runtime anomalies or issues that need to be addressed.
LA is calculated as the ratio of correctly predicted levels $L_c$ to the total log statements \(LA = \frac{L_c}{N_a}\).

\textbf{\textit{Metric 3: Average Level Distance (ALD):}} ALD further quantifies the \textit{severity deviation} of mispredicted levels from the reference level. We assign ordinal values to log levels: 
\textit{trace: 0, debug: 1, info: 2, warn: 3, error:4, fatal: 5}, and compute the absolute difference between predicted level $L_p^{(i)}$ and oracle level $L_a^{(i)}$ for each log. ALD is the mean deviation across all instances: $ ALD = \frac{1}{N_a} \sum_{i=1}^{N_a} \left| L_p^{(i)} - L_a^{(i)} \right| $.

\textit{\textbf{Metric 4: Message Accuracy (MA):}}
MA evaluates how accurately the predicted log messages match the oracle, which is essential for providing meaningful and relevant information during runtime. The content of log messages helps developers understand the system’s behavior, and inaccuracies in message generation can lead to confusion or missed insights during debugging. For Message Accuracy, we compare the predicted log messages to the actual messages in the source code. This metric is calculated by determining the number of log messages that are fully identical to the ground truth \(M_c\) and dividing it by the total number of log messages \(N_a\), yielding the accuracy value: \(MA = \frac{M_c}{N_a}\).

\textit{\textbf{Metric 5: Dynamic Expression Accuracy (DEA)}:} DEA evaluates whether generated logs preserve the structural integrity of runtime expressions, including both individual variables and composite logic (\eg ternary operators, arithmetic). For example, in the log template:\textit{("The server run on the ports, \{\}", args.status ? localPort : remotePort)}, the conditional expression \textit{(args.status ? localPort : remotePort)} is treated as a single semantic unit.
We aim to use this metric to ensure that the dynamic information recorded in logs remains consistent. This metric is calculated by taking the number of the exactly matched \(DP_c\) and dividing it by the total number of log statements \(N_a\) to obtain the accuracy value: \(DEA = \frac{DP_C}{N_a}\).

 \textit{\textbf{Metric 6: Static Text Similarity (STS) (with BLEU or ROUGE):}} 
 STS focuses on the static part of the log message. Unlike the dynamic variable, the static part always records the same information in log files, which will not vary due to the runtime behavior of software. For example, in the log message \textit{("The server is running, \{\}", status)}, "The server is running, \{\}" is regarded as the static part.
 Since this part primarily consists of natural language content, we use BLEU~\cite{Papineni2002BleuAM} and ROUGE~\cite{Lin2004ROUGEAP} metrics to evaluate the quality of the static text.
 In our implementation, we use the DM variant of BLEU~\cite{Chen2014ASC, Shi2021OnTE}, \ie the sentence-level BLEU without any smoothing method — coupled with ROUGE-L to holistically assess both lexical precision and long-sequence coherence. Specifically, ROUGE-L focuses on the longest common subsequence that effectively captures key operational patterns (e.g., error codes or API call chains) in multi-line logs, while BLEU's n-gram overlap measurement complements it by evaluating template fidelity at the token level.

 \input{insert/static_evaluation_task_formulation}

\subsection{Dynamic Evaluation}
\input{insert/dynamic_evaluation_workflow}
\label{sec:method:dynamic_evaluation}
\subsubsection{Task Formulation}
In dynamic evaluation, we provide the runtime perspective of evaluating the automatic logging tools. 
Two key evaluation metrics are employed: (1) compilability—ensuring the code with predicted log statements compiles without errors, and (2) log file similarity—measuring the alignment between log files generated by the predicted log statements and the oracle through textual similarity analysis. These criteria jointly validate both the functional correctness of log integration and the relevance of logged content.

As illustrated in Figure~\ref{fig:dynamic_evaluation_workflow}, the core innovation of our methodology lies in utilizing unit tests to emulate runtime log generation under controlled conditions. By executing these tests, we synthetically replicate software runtime states with minimal computational overhead, thereby enabling systematic collection of runtime logs that mirror real-world execution patterns for downstream analysis.
We define task formulation as a tuple \(<Code_{w/o\ \ LogStmt}, Code_{w/\ \ LogStmt}, Logs>\), where \(Code_{w/o\ \ LogStmt}\) is the code snippets without log statements covered by unit test, \(Code_{w/\ \ LogStmt}\) is the code snippets with predicted log statement, and \(Logs\) is the logs produced by those log statements in \(Code_{w/\ \ LogStmt}\) via unit tests. If the \(Code_{w/\ \ LogStmt}\) is not compiliable, \(Logs\) is null.

To build the dynamic evaluation pair based on our dataset, we begin by compiling the project to ensure all dependencies are resolved and the project is ready for execution. Next, we systematically identify all available unit tests within the project. For each unit test, we execute it individually while employing the Jacoco Plugin~\cite{Jacoco} to trace code coverage, specifically identifying whether the unit test covers any log statements in the codebase. Simultaneously, we use the SureFire Plugin~\cite{surefire} to capture the logs generated during the execution of the unit tests.
By correlating Jacoco coverage data with SureFire logs, we can match specific code snippets containing log statements to the corresponding unit tests that cover them, along with the runtime logs they generate.
We finally built 2,238 instances for dynamic evaluation.

\subsubsection{Metrics}
\label{sect:method_dynamic_metrics}
We propose four dynamic evaluation metrics to assess the similarity of logs printed by the generated and reference log statements in runtime execution.

\textit{\textbf{Metric 7: Compilation Success Rate (CSR):}} CSR measures the compilability of the predicted log statements. Due to issues such as undefined variables in the predictions or missing/outdated dependencies in the project environment, not all code snippets with predicted log statements can successfully compile.  We recorded the successfully compiled code snippet number as \(C_s\) and the evaluation pair number as \(C_a\). The metric is then calculated as: \(CSR = \frac{C_s}{C_a}\).
 
 \textit{\textbf{Metric 8: Log File Similarity (LFS) (with COS, BLEU, ROUGE):}}
 LFS evaluates how closely logs generated by predicted statements match those produced by ground truth statements.
 We remove log headers (e.g., timestamps) and only assess the main message body of each log entry to eliminate unnecessary differences.
 For a comprehensive assessment, we apply multiple similarity measures, including Cosine Similarity (COS)~\cite{Salton1975AVS}, BLEU~\cite{Papineni2002BleuAM}, and ROUGE~\cite{Lin2004ROUGEAP}. Cosine Similarity, commonly used in text analysis, calculates the cosine of the angle between two TF-IDF~\cite{SprckJones2021ASI} vectors, yielding 1 for identical vectors and 0 for orthogonal ones. Using TF-IDF, we down-weight frequent terms, emphasizing distinctive content in logs. This method effectively captures the similarity between meaningful log content, filtering out redundant information for a more accurate relevance measure.
ROUGE, on the other hand, focuses on recall by comparing n-grams between the predicted and reference logs. It evaluates how much of the reference content is preserved in the prediction. The most commonly used variant is ROUGE-N, which calculates the overlap of n-grams between two texts. 


 \textit{\textbf{Metric 9: False Positive Log Generation Rate (FPLR):}}
FPLR measures the proportion of predicted log statements that generate logs during unit test execution when the ground truth log statements would not have produced any logs. It helps assess whether the predicted log statements introduce unnecessary or redundant logs in scenarios where no log should be generated. The number of false positive instances is recorded as \(FP\), and the total number of predictions is \(P\). the metric is calculated as: \(FRLR=\frac{FP}{P}\).

 \textit{\textbf{Metric 10: False Negative Log Generation Rate (FNLR):}}
FNLR evaluates the proportion of predicted log statements that fail to generate logs during unit test execution when the ground truth log statements should have produced logs. It highlights instances where the predicted logs miss important events or information. The number of false negative instances is recorded as \(FN\), and the total number of predictions is \(P\). The metric is calculated as: \(FNLR = \frac{FN}{P}\).


%% file: insert/table_dataset_projects.tex
\begin{table}[htbp]
\caption{Details of \methodname dataset.}
\label{table:dataset}
\centering
\scalebox{0.8}{
\begin{tabular}{cccc}
    \toprule
    \textbf{Dataset} & \textbf{Domain}  & \textbf{\# Code {w/ LogStmt}} & \textbf{\# LogStmt}  \\
    \midrule
    Dbeaver & Database Management & 1,178 & 1,707 \\
    Dolphinscheduler & Task Scheduling & 883 & 1,855 \\
    Doris & High Performance Database & 1,635 & 2,860 \\
    Flink & Data Processing & 1,916 & 3,150 \\
    Hadoop & Distributed Storage & 8,351 & 14,987 \\
    Kafka & Messaging Systems & 1,667 & 3,308 \\
    Keycloak & Identity and Access Management & 491 & 926 \\
    Pulsar & Messaging Systems & 3,279 & 6,347 \\
    Thingsboard & IoT Platform & 1,544 & 2,588 \\
    Zookeeper & Distributed Coordination & 860 & 1,872 \\     
    \midrule
    \textbf{Total} & - & \textbf{21,804} & \textbf{39,600} \\
    \bottomrule
\end{tabular}
}
\end{table}

%% file: insert/static_evaluation_task_formulation.tex
\begin{figure}
    \centering
    \includegraphics[scale=0.7]
    {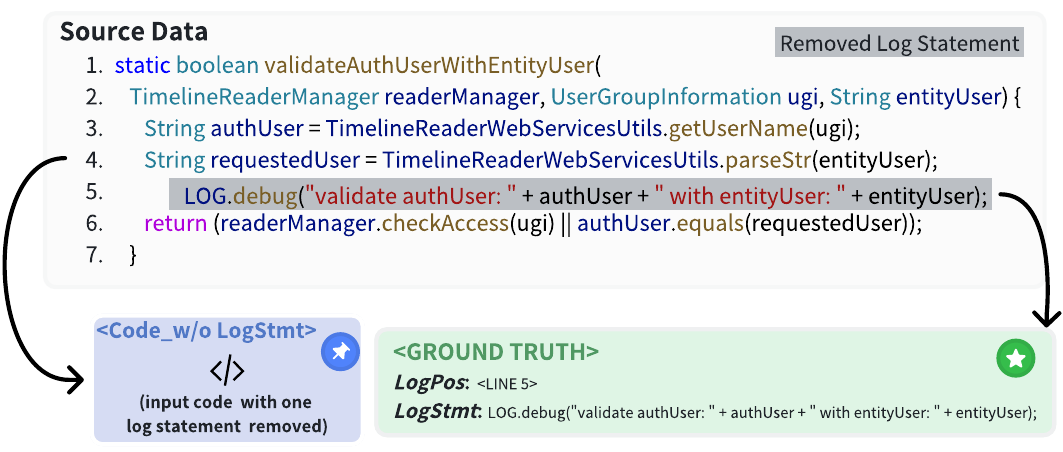}
    \caption{\(Code_{w/o\ \ LogStmt}\) indicates the code without one log statement, the \(LogPos\) means the position of this log statement, \(LogStmt\) is the log statement itself. Those three structure the evaluation tuple.}
\label{fig:static_evaluation_task_formulation}
\end{figure}

%% file: insert/dynamic_evaluation_workflow.tex
\begin{figure*}[t]
    \centering
    \includegraphics[width=1\textwidth]{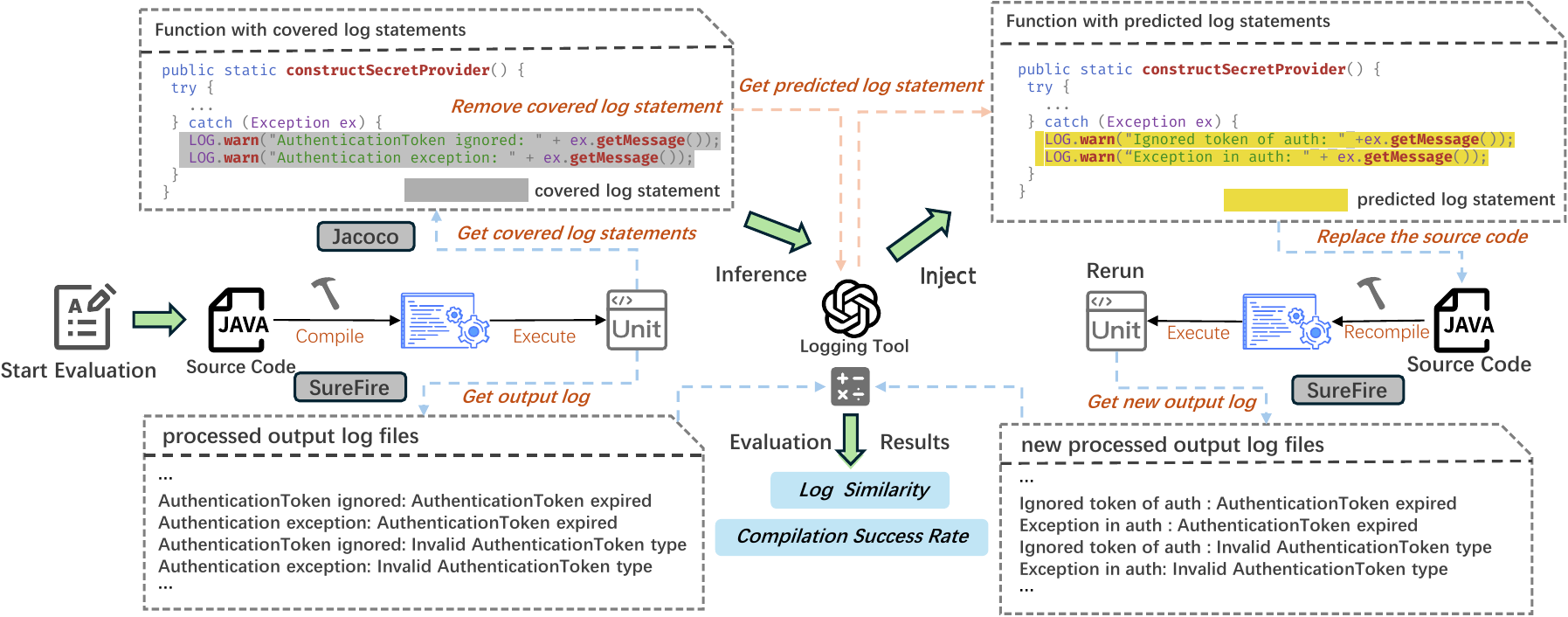}
    \caption{The general workflow of dynamic evaluation. First, compile the project and run the unit test to obtain ground truth logs. Then, replace log statements with predictions, re-run the test to generate new logs, and finally analyze the results.} 
    \label{fig:dynamic_evaluation_workflow}
\end{figure*}

%% file: sections/005_experiments.tex
\section{EXPERIMENTS}
\label{sec:eval}
In this section, we use \methodname to evaluate existing end-to-end automatic logging tools and analyze the evaluation results to find insights to guide the following work. From the perspective of empirical software engineering, we set three research questions:
\begin{itemize}
    \item \textbf{RQ1:} How similar are automatically generated log statements to human-written logging practices?
    \item \textbf{RQ2:} How compilable are the log statements generated by automatic logging tools?
    \item \textbf{RQ3:} How consistent are runtime logs from auto-logged programs with oracle-generated logs?
\end{itemize}

    Specifically, RQ1 aims to evaluate the performance of the tools in generating log statements that closely resemble the ground truth. RQ2 focuses on assessing the compilability of the predicted log statements, determining whether these predictions can be seamlessly integrated into the code without causing compilation errors. Finally, RQ3 evaluates the runtime logs produced by the generated log statements. We examine the similarity between the predicted logs and the expected logs. We begin by introducing the automatic logging tools selected for evaluation. Then we provide a detailed analysis of these tools by answering the three research questions. This comprehensive evaluation allows us to draw insights into the strengths and limitations of current approaches for automatic logging.

\subsection{Automatic Logging Tools}
Automatic logging is a hot topic, leading to the development of many tools for determining specific parts of log statements and end-to-end automatic logging tools in recent years. In this paper, we focus on end-to-end logging tools. We reached out to the authors of popular end-to-end automatic logging tools for assistance in rebuilding these tools. Because SCLogger~\cite{Li2024GoSC} is still under construction,
we ultimately selected four methods for evaluation: \textit{LANCE~\cite{Mastropaolo2022UsingDL}, LEONID~\cite{Mastropaolo2023LogSG}, FastLog~\cite{xie2024fastlog}, UniLog~\cite{Xu2024UniLogAL}}. We detailed the method in the following.

\textit{\textbf{LANCE:}} \(LANCE\)~\cite{Mastropaolo2022UsingDL} is the first model designed to generate and insert complete log statements in code. It takes a method requiring a log statement and outputs a meaningful log message with an appropriate logging level in the correct position. Built on the Text-To-Text Transfer Transformer (T5) model~\cite{Raffel2019ExploringTL}, \(LANCE\) is trained specifically for injecting proper logging statements.

\textit{\textbf{LEONID}:} \(LEONID\)~\cite{Mastropaolo2023LogSG} is the updated version of \(LANCE\). With a combination of DL and Information Retrieval (IR), \(LEONID\) achieved a better performance. \(LEONID\) provided two versions, \(LEONID_S\) is for single log statement generation, and \(LEONID_M\) is for multiple log statements generation. Since \(LEONID_M\) can generate more than one log statement at a time, it introduces ambiguity in determining the correct correspondence between the generated and expected log statements when more than one log statements are generated by static evaluation~\cite{Mastropaolo2023LogSG}. Therefore, we only applied \(LEONID_S\) for static evaluation.

\textit{\textbf{UniLog:}} \(UniLog\)~\cite{Xu2024UniLogAL} is the first attempt to adapt Warm-up and In-context-learning strategy to enhance the model's ability to generate log statements.
Due to limitations in assessing the original \textit{UniLog}, we reproduced it using two backbone models: CodeLlama-7B~\cite{Rozire2023CodeLO} and DeepSeek-V3~\cite{DeepSeekAI2024DeepSeekV3TR}. We applied the warmup process exclusively to the CodeLlama-7B backbone model while employing the ICL strategy to construct prompts for both models.
The data are sourced from \textit{LANCE}~\cite{Mastropaolo2022UsingDL} to warm up and generate ICL content. The effectiveness of In-Context Learning often depends on whether the examples are in-distribution or out-of-distribution relative to the evaluation data. Since \textit{LANCE’s} data distribution differs from our evaluation data, this may affect \textit{UniLog’s} performance. We will use \(\bm{UniLog_{{cl}}}\) to represent the version based on CodeLlama-7B, \(\bm{UniLog_{{ds}}}\) to represent the version based on DeepSeek-V3.

\textit{\textbf{FastLog:}} \textit{FastLog}~\cite{xie2024fastlog} defines the logging task in two steps: finding the position and generating and inserting a complete log statement into the source code. This approach avoids rewriting the source code, a key limitation of \(LANCE\). They utilized PLBART~\cite{Ahmad2021UnifiedPF} as the base model to fine-tune two separate models: one for predicting insertion position, the other for generating log statements. With the heuristic rule, log statements only appear after certain special characters, \textit{FastLog} enhances efficiency while maintaining accuracy in generating log statements.

\input{insert/table_static_evaluation_results}
\subsection{RQ1: How similar are automatically generated log statements to human-written logging practices?}
\label{rq1}
\input{insert/table_static_length_compare}
We evaluated four automatic logging tools using \methodname, leveraging 39,600 static evaluation pairs.
The evaluation results are presented in Table~\ref{table:static_evaluation_results}. 
All logging tools show average accuracy drops of 37.49\%, 23.43\%, and 15.80\% in predicting log position, level, and message compared to their reported results. Most strikingly, \(LEONID_S\) collapses from 31.55 → 1.96 in MA (\(\downarrow93.8\%\)) and \(UniLog_{cl}\) plummets from 76.90 → 23.49 in PA (\(\downarrow69.5\%\)). Those drops expose severe overfitting to narrow, synthetic datasets in prior studies.
When evaluated on \methodname, state-of-the-art tools achieved limited scores of 18.00 (DEA), 20.14 (BLEU-4), and 29.88 (ROUGE-L) in STS. These results highlight persistent limitations in generating semantically coherent natural language descriptions and reliably identifying contextually relevant dynamic content for logging.



To investigate the performance discrepancy relative to prior studies, we partitioned the dataset into two subsets: instances with sub-512 token lengths and those exceeding 512 tokens. This division enables us to determine whether instance length contributes to the performance discrepancies observed among the logging tools, providing further insight into how different data characteristics influence tool effectiveness.
After dividing the data, we obtained two groups: long data containing 8,925 instances and short data with 30,675 instances. The evaluation results in Table~\ref{table:static_len_compare} show a clear difference in tool performance based on the length of the instances.
Notably, \textit{LANCE} and \textit{LEONID} struggled with instances longer than 512 tokens, failing to generate syntax-correct code and, in some cases, producing incomplete code. This explains why their scores for these cases are reported as zero, highlighting the input length limitations of both tools and their inability to handle longer, more complex instances effectively.
\(UniLog_{cl}\) and \textit{FastLog} show a considerable drop in PA and DEA when handling longer data, indicating that they struggle to predict log positions and select dynamic variables in more complex instances. \(UniLog_{ds}\) exhibits a significant decline in positional accuracy but only marginal degradation in dynamic variable identification. Notably, it demonstrates stronger performance on metrics such as MA and STS when processing longer input sequences. This observation implies that enhancing a model’s architectural foundation and providing richer contextual data can improve its capacity to generate syntactically and semantically coherent logging statements through greater contextual awareness.

Although \(UniLog_{ds}\) is equipped with a more powerful backbone, it still struggles to choose the appropriate position for the log statement. As widely demonstrated that the LLMs are not good at counting numbers~\cite{Ahn2024LargeLM}, making LLM generate the exact line number might not be a wise choice. Furthermore, by simply analyzing the control flow graph of the code, we might be able to exclude positions where logging is not feasible to leave fewer choices.
\input{insert/pic_projects_diversity}

To evaluate tool stability, we analyzed performance consistency across individual projects. As shown in Figure~\ref{fig:project_diveristy}, tool performance exhibits significant variability across projects. For instance, in Flink, the state-of-the-art tool achieves an MA score of 13.68, contrasting sharply with the overall project average of 6.93. Similarly, in Keycloak, the state-of-the-art DEA reaches 43.84 compared to the aggregate average of 16.60. These stark discrepancies underscore inconsistencies in the reliability and generalizability of logging tools when applied to projects with distinct requirements and contexts.

\begin{tcolorbox}
    \textbf{Answer to RQ1.} 
    All logging tools show average accuracy drops of 37.49\%, 23.43\%, and 15.80\% in
predicting log position, level, and message compared to their reported results.
    All logging tools exhibit persistent limitations in generating semantically coherent log descriptions and reliably identifying contextually relevant dynamic variables. Moreover, they fail to achieve consistent performance reliability across diverse logging requirements and project-specific contexts.
\end{tcolorbox}

\input{insert/table_compiled_failed}

\subsection{RQ2: How compilable are the log statements generated by automatic logging tools?}
\label{rq2}
To evaluate the tools’ ability to generate compilable log statements, we replace existing log statements with predicted ones and recompile the project to check for successful compilation. As shown in Table~\ref{table:compile_failed_rate},
the best-performing tool, \textit{FastLog}, achieves a 79.9\% Compilation Success Rate, followed by \(UniLog_cl\) at 70.3\%, \(UniLog_ds\) at 60.2\%, \textit{LANCE} at 49.4\%, \(LEONID_M\) and \(LEONID_S\) at 25.0\% and 16.4\%, respectively.
The results reveal a key limitation of \textit{LANCE} and \textit{LEONID} that they regenerate the entire code snippet, which increases the risk of unintended code changes and can potentially lead to compilation errors. In contrast, \textit{FastLog} and \textit{UniLog} focus solely on generating the new log statement, minimizing the risk of errors by limiting codebase modifications. Although \textit{FastLog} is one of the best tools available, it still leads to significant instances (20.1\%) of compile failure. 



To understand the reason for the compilation failures reason, we conducted a manual review of the compilation failures to identify the specific causes. 
We randomly selected 100 failed instances from the best tool, \textit{FastLog}, and the first two authors cross-checked the causes. The analysis results are presented in Table~\ref{table:compile_failed_reason}. The most common failure, occurring in 56 instances, is due to using the wrong logging name~(\ie \textit{Logger}, \textit{LOG}), indicating that incorrect or non-existent log functions are being invoked. Using undefined methods accounts for 21 failures, followed by using undefined variables with 15 failures. Less frequent issues include incompatible types (3), and unreachable statements (1). Others mean generating the wrong syntax code, which we will not analyze.

The majority of failures (totaling 92 instances) involve undefined references to methods (21 instances), variables (15 instances), or logging names (56 instances). These undefined reference failures are primarily due to the limited context provided by existing function-level logging tools, lacking critical details on valid variables, methods, libraries, and packages relevant to the target function. Current tools are designed for function-level input, highlighting the need for logging tools to integrate better context awareness and validation checks to ensure compatibility with the existing codebase. The less frequent errors are also noticeable. For instance, the issue of unreachable statements points to a weakness in that tools lack an understanding of the code’s control flow, resulting in log statements placed in non-executable paths. Collectively, these errors highlight the need for automated logging tools to adopt enhanced context-aware strategies, improving the accuracy and contextual relevance of generated logs.


\begin{tcolorbox}
    \textbf{Answer to RQ2.} \textsc{FastLog} achieved the best performance in generating compilable log statements, yet over 20\% of the generated log statements still failed to be compiled. According to our analysis of compilation failure reasons, these failures primarily stem from a lack of critical contexts corresponding to the target function, e.g., valid variables, methods, libraries, packages, execution paths, and type information. To improve the reliability of automatic logging tools, it is crucial that they incorporate mechanisms to gather and utilize this additional context during the log statement generation process.
\end{tcolorbox}
 
\subsection{RQ3: How consistent are runtime logs from auto-logged programs with oracle-generated logs?}
\label{rq3}
To answer RQ3, we compare the semantic similarity of logs generated by predicted log statements and by the oracle.
It is important to highlight that, aside from \(LEONID_M\), all other tools operate under the strong assumption that the given code snippet requires exactly one log statement. 
To highlight the inappropriateness of this assumption, we only allow the tools one chance to predict the log statement, even in cases where multiple log statements are needed. 
Our experimental design emulates real-world logging scenarios by abandoning the assumption that single log statements universally suffice, thereby exposing a critical limitation in existing tools' capacity to determine contextually appropriate logging density for a given code context.

\input{insert/table_log_similarity}

Table~\ref{table:log_similarity} compares the semantic similarity between logs from source and predicted log statements.
The results indicate that logs from the predicted statements significantly deviate from the ground truth, with consistently low scores across metrics like Cosine Similarity, BLEU, and ROUGE.
 Limiting each code snippet to a single log statement often sharply compromises log similarity, especially when multiple statements are needed. This limitation is particularly clear when comparing \(LEONID_S\) and \(LEONID_M\). Although \(LEONID\) overall performs poorly, \(LEONID_M\) stands out as the only tool capable of generating multiple log statements, which enables it to outperform \(LEONID_S\) in similarity scores. This difference underscores the importance of tools being able to determine the appropriate number of log statements for accurate and effective logging, rather than assuming a single statement suffices.
\input{insert/table_FPLG_FNLG}

To understand the low semantic similarity scores, we examined the log generation process and found many instances where predicted logs were either redundant or missed key information present in the original logs. We quantify this issue using two metrics: FPLR and FNLR, as reported in Table~\ref{tab:table_FPLG_FNLG}. For example, \textit{FastLog} reports a 9.28\% FPLR, meaning logs record redundant information in 9.28\% of cases, and an 18.28\% FNLR, indicating expected information in logs was missing in 18.28\% of cases. We sampled 100 examples from FastLog, the state-of-the-art (SOTA) model, and manually analyzed the reasons for mismatches with the original logs. The results are presented in Table~\ref{tab:table_FPLG_FNLG_Reason}. We found that the primary reasons for failures in FNLR and FPLR differ significantly. For FNLR, the most common issue was the predicted log statements being beyond the execution path (35 cases), followed by lower verbosity levels (24 cases), and a smaller number caused by wrong code format (4 cases). In contrast, for FPLR, the main problem was higher verbosity levels (30 cases), with a few instances of log statements being beyond the execution path (3 cases). Overall, verbosity mismatches and execution path discrepancies were the dominant contributors, highlighting challenges in aligning predicted logs with actual logging requirements. The factors leading to FPLR and FNLR underscore a critical issue: while static metrics offer valuable insights into the quality of generated log statements, the actual logs are shaped by numerous contextual factors. Without a thorough understanding of the execution context, it is not possible to comprehensively evaluate the quality of log statements. Even minor discrepancies can cause significant deviations between generated logs and source logs. For instance, while the predicted log statement may capture the key information required to reflect system behavior, its effectiveness can be compromised if it is not positioned along a critical execution path or if its verbosity level is mismatched. In such cases, the log statement may fail to record essential information when key events occur. This issue majorly arises from the tools’ lack of awareness of verbosity thresholds and the control graph of the code, which limits their ability to adjust verbosity and determine appropriate log positions based on the context or execution requirements. These two limitations highlight that current logging tools lack the adaptability and context-awareness needed for effective real-world applications.

\begin{tcolorbox}
    \textbf{Answer to RQ3.} The best-predicted log statements by \textit{FastLog} achieve only 21.32\% cosine similarity with the original logs. Many predictions record redundant information, while others miss key details. The missing key information is primarily due to the prediction being placed beyond the execution path during important events, while setting a higher verbosity level in the log statements leads to redundancy. This result highlights that automatic logging tools still have significant room for improvement.
\end{tcolorbox}

%% file: insert/table_static_evaluation_results.tex
\begin{table*}[htbp]
\caption{Static evaluation on the complete logging task was conducted.
Note that lower \textbf{ALD} means better log level quality, and
\textbf{BLEU-4 \& ROUGE-L} refer to two kinds of \textbf{STS} metrics. The best results are highlighted.}
\label{table:static_evaluation_results}
\centering
\setlength{\tabcolsep}{8pt} 
\scalebox{1.0}{ 
\begin{tabular}{l|c|c|c|c|c|c|c}
    \toprule
    \multirow{1}{*}{\textbf{Method}} & \multirow{1}{*}{\textbf{PA}} & \multirow{1}{*}{\textbf{LA}} & \multirow{1}{*}{\textbf{MA}} & \multirow{1}{*}{\textbf{ALD}}& \multirow{1}{*}{\textbf{DEA}} & \multirow{1}{*}{\textbf{BLEU-4}}&\multirow{1}{*}{\textbf{ROUGE-L}}\\
    
    \midrule
    \textit{FastLog} & \(\textbf{58.39}\) & \(\textbf{62.15}\) & \(\textbf{6.93}\)& 0.63& \textbf{18.00} &\textbf{20.14}&29.32\\
    \(UniLog_{{ds}}\) & \(37.11\) & \(60.66\) & \(5.23\) & \textbf{0.61} & 16.20 &11.62&26.37\\
    \(UniLog_{{cl}}\) & \(23.49\) & \(50.97\) & \(2.71\) & 0.79 & 16.60 &8.79&\textbf{29.88}\\
    \textit{LANCE} & \(35.97\) & \(37.70\) & \(3.11\)  & 2.11 & 15.25 &6.7&15.64\\
    \(LEONID_S\) & \(11.26\) & \(17.90\) & \(1.96\) & 3.78 & 8.51 &3.45&6.69\\
    \bottomrule

\end{tabular}
}
\end{table*}

%% file: insert/table_static_length_compare.tex
\begin{table*}[htbp]
\caption{The performance of tools when facing different length input data. \textbf{Short} means data shorter than 512 tokens, \textbf{Long} means data longer than 512 tokens, and $\Delta$ means the difference in logging tools performance when facing the \textbf{Short} data and the \textbf{Long} data.}
\label{table:static_len_compare}
\centering
\setlength{\tabcolsep}{12pt} 
\scalebox{0.85}{ 
\begin{tabular}{l|c|cclcccc}
    \toprule
\textbf{Method} & \textbf{Dataset} & \textbf{PA} & \textbf{LA}  & \textbf{MA}&\textbf{ALD} & \textbf{DEA} & \textbf{BLEU-4} &\textbf{ROUGE-L}\\
    \midrule
\multirow{3}{*}{\textit{FastLog}} & Short& 62.27& 61.79& 6.81&0.61& 19.65& 19.71& 29.43\\
 & Long& 45.04& 63.37& 7.37&0.68& 12.35& 21.63& 28.9\\
 & $\Delta$ & $\downarrow 17.23$& $\uparrow 1.58$ & $\uparrow 1.58$&$\uparrow 0.56$& $\downarrow 7.30$& $\uparrow 1.92$& $\downarrow 0.47$\\
     \midrule
 \multirow{3}{*}{\(UniLog_d{}_s\)} & Short& 41.19& 59.00& 4.11&0.63& 16.21& 9.93&24.51\\
 & Long& 23.10& 66.38& 9.06&0.54& 16.19& 17.40&32.77\\
 & $\Delta$ & $\downarrow 18.09$& $\uparrow7.38$& $\uparrow4.95$&$\downarrow 0.09$& $\downarrow 0.03$& $\uparrow7.47$&$\uparrow8.26$\\
     \midrule
 \multirow{3}{*}{\(UniLog_c{}_l\)} & Short& 27.78& 51.07& 2.64&0.77& 17.53& 8.86&30.24\\
 & Long& 8.73& 50.59& 2.94&0.82& 13.39& 8.44&24.58\\
 & $\Delta$ & $\downarrow 19.05$ & $\downarrow 0.48$& $\uparrow 0.30$&$\uparrow 0.09$& $\downarrow 4.14$& $\downarrow 0.42$&$\downarrow 5.66$\\
     \midrule
 \multirow{3}{*}{\textit{LANCE}} & Short& 46.44& 48.67& 4.01&1.56& 19.69& 8.65&20.19\\
 & Long& 0& 0&0& 0& 0& 0&0\\
 & $\Delta$ & - & -  &-& - & - & -  &-\\
     \midrule
 \multirow{3}{*}{\(LEONID_S\)} & Short& 14.54& 23.11& 2.53&3.43& 10.99& 4.45&8.64\\
 & Long& 0& 0&0& 0&  0& 0&0\\
 & $\Delta$ & - & -  &-& - & - & -  &-\\
     \bottomrule
\end{tabular}
}
\end{table*}

%% file: insert/pic_projects_diversity.tex
    \begin{figure}[t]
    \centering
    \includegraphics[width=1.0\textwidth]{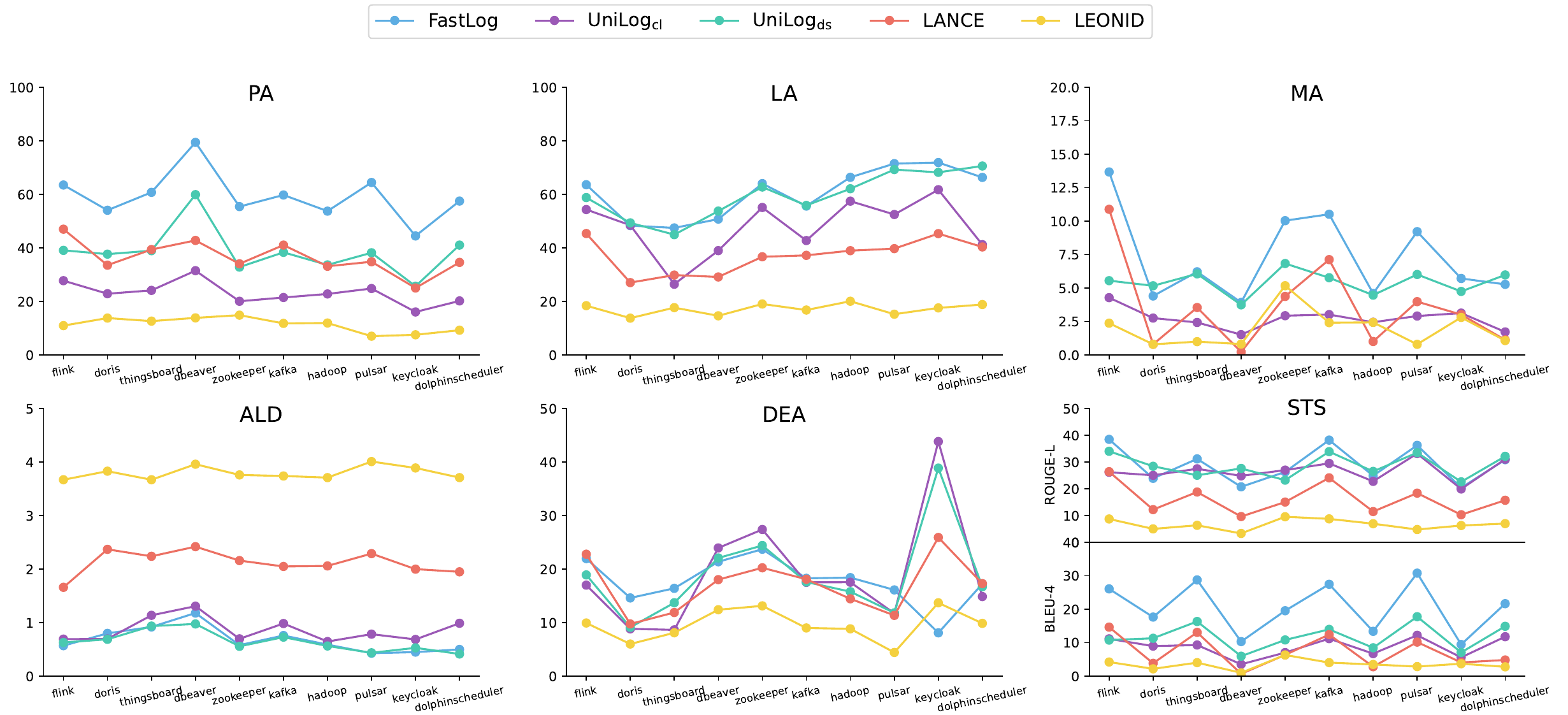}
    \caption{Performance of logging tools among different projects. The performance of each tool varies considerably across different projects and the trends of all methods across all projects generally remain consistent. }
         \label{fig:project_diveristy}
    \end{figure}

%% file: insert/table_compiled_failed.tex
\begin{table}[h]
    \centering
    \begin{minipage}{0.45\textwidth}
\caption{Compilation Failure Rates Across Methods.}
\label{table:compile_failed_rate}
\centering
\setlength\tabcolsep{12pt}  
\renewcommand{\arraystretch}{1.1}  
\scalebox{1.1} {  
\begin{tabular}{c|c}
    \toprule
\textbf{Methods} & \textbf{CSR} \\  
\midrule
\textit{\textbf{FastLog}} & \textbf{79.9\%} \\ 
\(UniLog_{{cl}}\) & 70.3\% \\ 
\(UniLog_{{ds}}\) & 60.2\% \\ 
\textit{LANCE} & 49.4\% \\ 
\(LEONID_M\) & 25.0\% \\
\(LEONID_S\) & 16.4\% \\ 
\bottomrule
\end{tabular}
}
    \end{minipage}
    \hfill
    \begin{minipage}{0.5\textwidth}
\caption{Compilation Failure Reasons Analysis.}
\label{table:compile_failed_reason}
\centering
\setlength\tabcolsep{12pt}  
\renewcommand{\arraystretch}{1.1}  
\scalebox{1} {  
\begin{tabular}{c|c}
    \toprule
\textbf{Failed Reason} & \textbf{\# Failed} \\     \midrule
Using Wrong Logging Name & 56 \\ 
Using Undefined Method & 21 \\ 
Using Undefined Variables & 15 \\ 
Incompatible Types & 3 \\
Unreachable Statements & 1 \\ 
Other & 4 \\     \midrule
Total & 100 \\
\bottomrule
\end{tabular}
}
    \end{minipage}
\end{table}

%% file: insert/table_log_similarity.tex
\begin{table*}[htbp]
\caption{ The semantic similarity between logs printed by the predicted log statements and the ground truth. }
\label{table:log_similarity}
\centering
\setlength{\tabcolsep}{6pt} 
\scalebox{0.9}{ 
\begin{tabular}{lcccccccccc}
    \toprule
    \textbf{Method} & \textbf{COS(\%)} & \textbf{BLEU-1} & \textbf{BLEU-2}& \textbf{BLEU-3}& \textbf{BLEU-4}& \textbf{ROUGE-1} & \textbf{ROUGE-2} & \textbf{ROUGE-L}\\
    \midrule
 \(FastLog\)       & \textbf{21.32}                 & \textbf{24.83}  &\ \textbf{19.62}  &\ \textbf{17.49}        & \textbf{15.96}            & \textbf{24.43}    &\ \textbf{18.74}      & \textbf{23.84}           \\ 
 \(UniLog_{{cl}}\)         & 17.38                  & 19.02    &\ 15.62  &\ 14.34        & 13.42          & 19.17     &\ 15.37       & 18.75           \\ 
\(UniLog_{{ds}}\) & 13.04                 & 14.67  &\ 12.72  &\ 12.02          & 11.56          & 14.00      &\ 12.23      & 13.87           \\ 
\(LANCE\)          & 9.93                   & 11.34   &\ 9.21  &\ 8.50         & 7.97           & 11.23      &\ 8.94      & 11.03           \\ 
 \(LEONID_M\)      & 7.19                  & 7.95    &\ 6.51  &\ 5.87         & 5.30           & 8.10      &\ 6.43       & 7.94            \\ 
 \(LEONID_S\)         & 4.45                 & 4.91     &\ 3.82  &\ 2.95        & 2.96           & 5.10        &\ 3.86     & 5.01            \\
\bottomrule
\end{tabular}
}
\end{table*}

%% file: insert/table_FPLG_FNLG.tex
\begin{table*}[htbp]
\centering
\begin{minipage}{0.48\textwidth}
\centering
\caption{False Positive and False Negative Logging Rates across Methods. \textbf{FPLR} means the predicted log statement record logs when it should not, and \textbf{FNLR} means the predicted log statement does not record logs when it should.}
\label{tab:table_FPLG_FNLG}
\setlength{\tabcolsep}{8pt} 
\begin{tabular}{lcc}
    \toprule
    \textbf{Method} & \textbf{FPLR} & \textbf{FNLR} \\
    \midrule
        \textit{FastLog} & 9.28\% & 18.28\% \\
        \(UniLog_{{cl}}\)  &	6.52\% &	30.59\% \\
        \(UniLog_{{ds}}\)  &	3.21\% &	 22.88\%\\
    \textit{LANCE}   & 5.71\% & 19.29\% \\
    \(LEONID_S\) &	8.15\% & 8.69\% \\
        \(LEONID_M\)  &	7.32\% & 11.60\% \\
    \bottomrule
\end{tabular}
\end{minipage}\hfill
\begin{minipage}{0.48\textwidth}
\centering
\caption{\textbf{FPLR} and \textbf{FNLR} Reason Analysis. For FNLR, the major reason is beyond the execution path, and for FPLR, the major reason is lower verbosity level.}
\label{tab:table_FPLG_FNLG_Reason}
\centering
\setlength\tabcolsep{12pt}  
\renewcommand{\arraystretch}{1.1}  
\scalebox{0.8} {  
\begin{tabular}{c|c|c}
    \toprule
\textbf{Situation} & \textbf{Reason} & \textbf{\# False} \\     \midrule
\multirow{3}{*}{\textbf{FNLR}}& Beyond Execution Path & 35  \\ 
& Lower Verbosity Level & 24\\ 
& Wrong Code Format & 4 \\ \midrule
\multirow{2}{*}{\textbf{FPLR}} & Higher Verbosity Level & 30 \\ 
& Beyond Execution Path & 3 \\    \midrule
Total & -  & 100 \\
\bottomrule
\end{tabular}
}
\end{minipage}
\end{table*}

%% file: sections/006_conclusion.tex
\section{Threats to Vadility}
\label{sec: vadility}
\begin{itemize}
    \item \textbf{Construct Validity}: The first threat comes from the use of two widely-accepted syntactic metrics in previous logging research (BLEU and ROUGE)~\cite{xie2024fastlog, Xu2024UniLogAL, He2018CharacterizingTN, Ding2023LoGenTextPlusIN}, as they may not fully capture the semantics of the log messages~\cite{Gros2020CodeTC, Roy2021ReassessingAE}. To mitigate this threat, we also adapt Cosine Similarity to evaluate the semantic quality of log messages.  Since all log headers are excluded, leaving log messages primarily in natural language, Cosine Similarity can effectively capture the log's semantic meaning~\cite{Salton1975AVS}.
    \item \textbf{Internal Validity}: The second threat to validity concerns reproducing the baseline. To minimize inconsistencies, we adapted the released models from previous work~\cite{xie2024fastlog, Mastropaolo2022UsingDL, Mastropaolo2023LogSG} and sought guidance from the authors of the closed-source model UniLog~\cite{Xu2024UniLogAL} to reproduce the tool under supervision.
    \item \textbf{External Validity}: 
    {The third threat comes from the potential data contamination risks in our dataset. 
    To mitigate it, we follow the methodology used by GPT-3~\cite{Brown2020LanguageMA} to assess data contamination severity, i.e., considering test samples with more than a 13-gram overlap with the original code samples as contaminated.
    Specifically, we perform a 13-gram overlap analysis between test and training samples of all evaluation methods. Results indicate a peak contamination rate of 1.24\% across all training datasets, empirically demonstrating no evidence of data leakage in our framework.} 
\end{itemize}

\section{Conclusion}
\label{sec: conclusion}

In this paper, we introduced \methodname, a comprehensive benchmark for automatic logging tools evaluation, featuring a diverse dataset from 10 established projects and a novel dynamic evaluation methodology. Our benchmark revealed significant limitations in existing tools through both static and dynamic perspectives: substantial accuracy drops in log position, level, and message prediction compared to reported results, high compilation failure rates (20.1\%-83.6\%), and low runtime log similarity (maximum 21.32\%). These findings highlight critical gaps between current tool capabilities and practical requirements. We believe \methodname establishes a solid foundation for advancing automatic logging research by providing standardized evaluation metrics and exposing areas requiring improvement.

\section{DATA AVAILABILITY}
All the code and data used in our study are publicly available on \url{https://github.com/shuaijiumei/logging-benchmark-scripts}.